\documentclass[a4paper,12pt]{article}

\pdfoutput=1

\usepackage[paper=letterpaper,margin=1in]{geometry}

\usepackage{amsmath,amssymb,amsfonts,epsfig,cite,setspace,bigstrut,longtable,array,tabularx,url,multirow}

\begin{document}

\newcommand{\Mobius}{M\"{o}bius }
\newcommand{\be}{\begin{equation}}
\newcommand{\ee}{\end{equation}}
\newcommand{\bea}{\begin{eqnarray}}
\newcommand{\eea}{\end{eqnarray}}

\def\half{{\scriptstyle {1\over 2}}}

\newcommand{\todo}[1]{{\bf ?????!!!! #1 ?????!!!!}\marginpar{$\Longleftarrow$}}
\newcommand{\fref}[1]{Figure~\ref{#1}}
\newcommand{\tref}[1]{Table~\ref{#1}}
\newcommand{\sref}[1]{\S~\ref{#1}}
\newcommand{\nn}{\nonumber}
\newcommand{\tr}{\mathop{\rm Tr}}
\newcommand{\ch}{\rm Ch}
\newcommand{\comment}[1]{}

\newcommand{\cM}{{\cal M}}
\newcommand{\cW}{{\cal W}}
\newcommand{\cN}{{\cal N}}
\newcommand{\cH}{{\cal H}}
\newcommand{\cK}{{\cal K}}
\newcommand{\cZ}{{\cal Z}}
\newcommand{\cO}{{\cal O}}
\newcommand{\cB}{{\cal B}}
\newcommand{\cC}{{\cal C}}
\newcommand{\cD}{{\cal D}}
\newcommand{\cE}{{\cal E}}
\newcommand{\cF}{{\cal F}}
\newcommand{\cR}{{\cal R}}
\newcommand{\cV}{{\cal V}}
\newcommand{\IA}{\mathbb{A}}
\newcommand{\IP}{\mathbb{P}}
\newcommand{\IQ}{\mathbb{Q}}
\newcommand{\IH}{\mathbb{H}}
\newcommand{\IR}{\mathbb{R}}
\newcommand{\IC}{\mathbb{C}}
\newcommand{\IF}{\mathbb{F}}
\newcommand{\IM}{\mathbb{M}}
\newcommand{\II}{\mathbb{I}}
\newcommand{\IZ}{\mathbb{Z}}
\newcommand{\re}{{\rm Re}}
\newcommand{\im}{{\rm Im}}
\newcommand{\sym}{{\rm Sym}}

\newcommand{\tmat}[1]{{\tiny \left(\begin{matrix} #1 \end{matrix}\right)}}
\newcommand{\mat}[1]{\left(\begin{matrix} #1 \end{matrix}\right)}
\newcommand{\diff}[2]{\frac{\partial #1}{\partial #2}}
\newcommand{\gen}[1]{\langle #1 \rangle}
\newcommand{\ket}[1]{| #1 \rangle}
\newcommand{\jacobi}[2]{\left(\frac{#1}{#2}\right)}

\newcommand{\drawsquare}[2]{\hbox{%
\rule{#2pt}{#1pt}\hskip-#2pt%  left vertical
\rule{#1pt}{#2pt}\hskip-#1pt%  lower horizontal
\rule[#1pt]{#1pt}{#2pt}}\rule[#1pt]{#2pt}{#2pt}\hskip-#2pt%  upper horizontal
\rule{#2pt}{#1pt}}% right vertical
\newcommand{\fund}{\raisebox{-.5pt}{\drawsquare{6.5}{0.4}}}
\newcommand{\antifund}{\overline{\fund}}

\newtheorem{theorem}{\bf THEOREM}
\def\thetheorem{\thesection.\arabic{theorem}}
\newtheorem{proposition}{\bf PROPOSITION}
\def\thetheorem{\thesection.\arabic{proposition}}
\newtheorem{observation}{\bf OBSERVATION}
\def\thetheorem{\thesection.\arabic{observation}}

\def\theequation{\thesection.\arabic{equation}}
\newcommand{\setall}{\setcounter{equation}{0}
        \setcounter{theorem}{0}}
\newcommand{\setequation}{\setcounter{equation}{0}}
\renewcommand{\thefootnote}{\fnsymbol{footnote}}

~\\
\vskip 1cm

\thispagestyle{empty}

\begin{center}
{\Huge \bf The Scattering Variety}
\end{center}
\medskip

\vspace{.4cm}
\centerline{
{\large Yang-Hui He}$^{1,2,3}$  ,
{\large Cyril Matti}$^1$ \&
{\large Chuang Sun}$^3$
\footnote{
hey@maths.ox.ac.uk; \ 
cyril.matti.1@city.ac.uk; \
chuang.sun@physics.ox.ac.uk
}
}
\vspace*{3.0ex}

\begin{center}
{\it
{\small
{${}^{1}$ 
Department of Mathematics, City University, London, EC1V 0HB, UK;\\
${}^{2}$
School of Physics, NanKai University, Tianjin, 300071, P.R.~China;\\
${}^3$
Rudolf Peierls Centre for Theoretical Physics, Oxford University, OX1 3NP, U.K.
}
}}
\end{center}

\vspace*{4.0ex}
\centerline{\textbf{Abstract}} \bigskip
The so-called Scattering Equations which govern the kinematics of the scattering of massless particles in arbitrary dimensions have recently been cast into a system of homogeneous polynomials.
We study these as affine and projective geometries which we call Scattering Varieties by analyzing such properties as Hilbert series, Euler characteristic and singularities. Interestingly, we find structures such as affine Calabi-Yau threefolds as well as singular K3 and Fano varieties.

\newpage

\setcounter{page}{1}

\tableofcontents

%\newpage

%%%%%%%%%%%%%%%%%%%%%%%%%%%%%%%%%%%%%%%%%
%%%%%%%%%%%%%%%%%%%%%%%%%%%%%%%%%%%%%%%%%
\section{Introduction}\setall

Recently, a programme was launched  to study the kinematics of the scattering of massless particles in arbitrary dimensions \cite{Cachazo:2013iaa,Cachazo:2013gna,Cachazo:2013hca,Cachazo:2013iea,Dolan:2013isa,Dolan:2014ega,Weinzierl:2014vwa, Naculich:2014naa, Naculich:2014rta}.
The initial motivation was the reduction of the tree-level S-matrix
of a wide range of theories as an integral over the moduli space of maps from the $N$-punctured sphere into the null light-cone in momentum space \cite{Witten:2003nn,Roiban:2004yf,Cachazo:2012da,Cachazo:2012kg,Huang:2012vt,Spradlin:2009qr}.

Suppose we are given $N$ null-vectors in $D$-dimensions corresponding to the momenta of $N$ massless particles:
\begin{equation}
\{k_1^\mu, \ k_2^\mu, \ \ldots, \ k_N^\mu \}\ , \qquad
k^2 = 0 \ .
\end{equation} 
The map that retrieves the scattering data 
from the $N$-punctured sphere with complex coordinate $z$ and punctures $\sigma_a$ is obtained from \cite{Cachazo:2013iaa, Cachazo:2013gna}:
\begin{equation}\label{map}
\sigma_a \mapsto k_a^\mu =
\frac{1}{2\pi i} \oint_{|z - \sigma_a| = \epsilon} dz
\frac{P^\mu(z)}{\prod\limits_{b=1}^N (z-\sigma_b)} \ ,
\end{equation}
where $P^\mu(z)$ is a collection of $D$ (indexed by $\mu$)
degree $N-2$ polynomials.

The massless condition on $k^\mu$ clearly translates to that $P(z)^2 = 0$. Moreover, the polynomial $P$ must be a null vector for all values of $z$, therefore, taking the derivative gives $P(z) \cdot \partial_zP(z) = 0$. This condition evaluated on the $N$ punctures locations $\sigma_a$, together with~\eqref{map}, was shown to be equivalent to:
\begin{equation}\label{scatter}
\sum\limits_{b \ne a} \frac{k_a \cdot k_b}{z_a - z_b} = 0 \ ,
\quad
a,b = 1, 2, \ldots, N \ .
\end{equation}
See~\cite{Cachazo:2013iaa, Cachazo:2013gna} for details.
Hence, the above equations govern the kinematics of our problem and were dubbed the {\bf scattering equations}.

Our main focus will be the study of these equations.
In fact, the nice paper \cite{Dolan:2014ega} has just reduced them to a system of homogeneous polynomials; it is with the geometry of this system, to which {\it cit.~ibid.}~already alluded, that we shall be chiefly concerned.

Throughout this work, we will adhere to the nomenclature of \cite{Dolan:2014ega}. We label the $N$ particles with the set of indices,
\begin{equation}
A = \{1,2,3, \ldots, N\} \ ,
\end{equation}
for the momenta $k_{a \in A}$ and introduce complex variables $z_{a \in A}$.
Then, we consider all subsets $S$ of $A$ with $m$ elements and define,
\begin{equation}
k_S := \sum\limits_{a \in S} k_a \ ; \mbox{ and }
z_S := \prod\limits_{b \in S} z_b \ , \mbox{ such that }
S \subset A \ .
\end{equation}

The insight of \cite{Dolan:2014ega} is that the scattering equations \eqref{scatter} are equivalent to the polynomial systems:
\begin{equation}\label{scatterhomo}
\tilde h_m = 0 \ , \quad \mbox{with} \quad
\tilde h_m := \sum_{S\subset A\atop |S|=m} k_S^2 z_S
 \ , \quad 2 \le m \le N-2 \ ,
\end{equation}
for null conserved momenta, meaning that the number of independent parameters $k$ additionally satisfy the constraints,
\be\label{kcons}
 k_a\cdot k_a=0 \ , \quad \mbox{and} \quad \sum\limits_a k_a = 0 \ .
\ee
Thus defined, $V_N : = \langle \tilde h_m \rangle$ can be seen as a polynomial ideal consisting of a set of $N-2-2+1 = N-3$ polynomials, each homogeneous of degree $m$ in the $N$ complex variables $z_a$, and such that $k_a$ are parameters obeying~\eqref{kcons}. Therefore, the quotient ring $M_N:= \mathbb{C}[{z_1},...,{z_{N}}]/V_N$ defines an affine algebraic variety $\cV_N$ in $\IC^N$, which can be seen as an affine cone over a compact projective variety $\cB$ in $\IC\IP^{N-1}$.

The homogeneous polynomials \eqref{scatterhomo} have the peculiar characteristic that when all coefficients $k_S^2\ne 0$ then the values of $z_a$ are distinct for all $a\in A$. This is reminiscent of the way the scattering equations factorize when one of the Mandelstam variable vanishes and has been demonstrated in~\cite{Dolan:2014ega}. Henceforth, we will impose the condition $k_S^2\ne 0$ to hold for the varieties we study.
It should also be pointed out that each polynomial in $V$, despite being of degree exceeding or equal to $2$, is linear in each variable $z_a$ considered separately.
Hence, we have an ideal which is square-free and multi-linear in all the coordinates.

Moreover, the scattering equations have the nice property to be M\"{o}bius invariant and \cite{Dolan:2014ega} demonstrated that the equivalent set of polynomials \eqref{scatterhomo} form an irreducible representation of the \Mobius algebra.
In this work, we draw on the insight of \cite{Dolan:2014ega} and explore the nature of general algebraic varieties $\cV_\varphi$ defined from polynomial systems resulting from irreducible representations of the \Mobius algebra. We also study the corresponding varieties $\cV_\varphi^*$ that result when \Mobius transformations are used to partially fixed two of the variety variables.

We find that they are all affine Calabi-Yau manifolds and, for the special case of the varieties $\cV_N$ corresponding to the scattering equations, the Hilbert series coefficients in the second kind have the pattern of the {\it Mahonian triangles}. The corresponding varieties are all singular and consist of a Conifold, a K3 surface and Fano varieties for the number of scattering particles $4,5,6$ and $7$ respectively. In addition, for the varieties $\cV_\varphi$ defined from irreducible \Mobius representation of spin $\half N-3$, we find that the Hilbert series coefficients correspond to the {\it cyclotomic polynomials}. We also demonstrate that the physical constraints~\eqref{kcons} on the polynomial coefficients imply for all the varieties $\cV_\varphi$ to be singular in at least one point.

The plan of the paper is as follows.
In the following section, we review irreducible representations of the \Mobius algebra and their relation to the scattering equations. We then present our results in the subsequent section, describing the nature of the geometries constructed. The last section presents our conclusions.

%%%%%%%%%%%%%%%%%%%%%%%%%%%%%%%%%%%%%%%%%
%%%%%%%%%%%%%%%%%%%%%%%%%%%%%%%%%%%%%%%%%
\section{Polynomial Systems}\setall

This section reviews the relation of the scattering equations to irreducible representations of the \Mobius algebra that has been presented in \cite{Dolan:2014ega}. We will not present the derivations in full length but will summarise the most important conclusions. We do so to set our notations\footnote{We will mainly follow the notations from \cite{Dolan:2014ega} with small variations for the sake clarity.} and motivate the geometrical considerations that are presented in the subsequent section.

\subsection{M\"{o}bius algebra}

The scattering equations \eqref{scatter} are invariant under \Mobius transformations and, consequently, so are \eqref{scatterhomo}. This can be seen as follows.
For complex numbers $\alpha, \beta, \gamma$ and $\delta$ satisfying $\alpha\delta-\beta\gamma \neq 0$, we can define the \Mobius transformation,
\be
 z_a \rightarrow \zeta_a = \frac{\alpha z_a + \beta}{\gamma z_a + \delta} \ , \quad a \in A \ .
\ee
Then $\zeta_a$ are also solutions of \eqref{scatter} when $z_a$ are solutions themselves. In fact, \cite{Dolan:2014ega} showed that the polynomials \eqref{scatterhomo} form a basis of an irreducible $(N-3)$-dimensional representation of the \Mobius algebra $\mathfrak{sl}_2(\IC)$, in a way which we now quickly summarise.

Let us consider the following operators acting on $\IC[z_a]$, the ring of polynomials in $z_a$ with $a\in A$,
\bea
\nn
L_{0} &=& -\sum_{a\in A}z_a\frac{\partial}{\partial z_a} + \frac{N}{2}\ , \\
\nn
L_{1} &=& \phantom{-}\sum_{a\in A}z_a - z_a^2\frac{\partial}{\partial z_a}\ , \\
\label{L}
L_{-1} &=& -\sum_{a\in A}\frac{\partial}{\partial z_a}\ .
\eea
It is straightforward to verify that these operators satisfy the $\mathfrak{sl}_2(\IC)$ commutation relations ,
\be\label{comm}
 \left[L_1, L_{-1}\right]=2L_0 \ , \quad \left[L_0, L_{\pm 1}\right]=\mp L_{\pm 1} \ ,
\ee
and, therefore, generate the \Mobius group $PSL(2,\IC)$. In fact, the ring of polynomials in $z_a, a\in A$ defines an infinite-dimensional representation space for the \Mobius transformations. Moreover, the subspace of \Mobius invariant polynomials provides a graded finite-dimensional representation space that decomposes into irreducible subspaces \cite{Dolan:2014ega}.

Acting on the polynomials $\tilde h_m$, the above operators act as raising and lowering (creation and annihilation) operators for the index $m$. In fact, they generate the set of all polynomials $\tilde h_m$ starting from the lowest degree polynomial $\tilde h_2$. Indeed, repeated action of $L_1$ generate $\tilde h_m$ for $m > 2$,
\be\label{hgeneration}
  (L_{1})^{r}\tilde h_2=r!\ \tilde h_{2+r} \ .
\ee
Moreover, the polynomials $\tilde h_m$ have the property that $L_{-1}\tilde h_2=0$ and $L_1\tilde h_{N-2}=0$, hence the index $m$ takes values $2 \le m \le N-2$, making $\tilde h_m$ closed under the \Mobius representation~\eqref{comm}. The corresponding quadratic Casimir for this representation is given by
\be
 L^2:=L_0^2 - \half L_1L_{-1} - \half L_{-1}L_1 \ ,
\ee
and, acting on $\tilde h_m$, takes value $(\half -2)(\half N -1)$. Therefore, the representation spanned by the $\tilde h_m$ polynomials form an irreducible $(N-3)$-dimensional representation of \Mobius spin $\half N-2$.

%%%%%%%%%%%%%%%%%%%%%%%%%%%%%%%%%%%%%%%%%
\subsection{Algebraic varieties}\label{algvar}

Drawing on the above observations for $\tilde h_m$, we can consider the other, more general, irreducible subspaces that arise from different lowest degree polynomials, as described in~\cite{Dolan:2014ega}. The eigenvalues of the $L_0$ operator are the largest for these lowest degree polynomials and they are therefore referred to as the highest weight polynomials.

A generic \Mobius invariant polynomial must be linear in each of its variables taken separately. Therefore, a generic such polynomial $\varphi_m \in \IC[z_a]$ of degree $m$ can be written in the following way,
\be
 \varphi_m := \sum_{S\subset A\atop |S|=m} \lambda_S z_S \ ,
\ee
where $\lambda_S$ are tensors with indices in $S$ and vanish if any two indices are equal. We should also note that the tensor indices $\lambda_{i_1\dots i_m}$ for $S=\{i_1\dots i_m\}$ are totally symmetric in $i_1\dots i_m$ and, in addition, we impose the constraint that $\lambda_S\ne 0$ for all subsets $S$~\footnote{The case of $\lambda_S = 0$ corresponds to the scattering of particles when some of the Mandelstam variables vanish, which we will not consider.}.

In order to define an irreducible representation of the \Mobius algebra, we must select a highest weight polynomial $\varphi_n$. We will reserve the notation $n$ for indices of highest weight polynomials, whereas the full series of polynomials will be noted with an index $m$, hence $n=\mbox{min}(m)$. Since the representation must close under the representation~\eqref{comm}, we must impose the condition,
\be
 L_{-1} \varphi_n =0 \ .
\ee
This translates onto the tensor coefficients
\be\label{HW}
\sum_{S\subset A\ |\ a\in S,\atop |S|=n  }\lambda_S = 0 \ , \quad \mbox{for each}\quad a\in A \ .
\ee
We will henceforth refer to these constraints as the {\bf highest weight} conditions. It has been shown in~\cite{Dolan:2014ega} that they are sufficient conditions to generate an irreducible representation of the \Mobius algebra acting on $\varphi_n$ with $L_1$ repeatedly,
\be
(L_1)^r\varphi_n = r!\ \varphi_{n+r} \ .
\ee
This series terminates at $L_1\varphi_{N-n}=0$ and thus the index $m$ range is $n\le m\le N-n$, giving a set of $N-2n+1$ polynomials. Generating the representation this way will imply some structure on the tensor coefficients $\lambda_S$. We can write for the full set of polynomials,
\be\label{poly}
 \varphi_m=\sum_{S \subset A \atop |S| = m} \lambda^{(n)}_S z_S \ , \quad \mbox{with} \quad \lambda^{(n)}_S=\sum_{U \subset S \atop |U| = n}\lambda_U \ ,
\ee
where $\lambda^{(n)}_S$ are defined to be the tensor coefficient of the highest weight polynomial $\varphi_n$ and satisfy the conditions~\eqref{HW}. The quadratic Casimir operator takes value
\be
 L^2\varphi_m=(\half N-n)(\half N-n+1)\varphi_m \ .
\ee
Hence, any highest weight polynomial $\varphi_n$ will generate an irreducible representation of the \Mobius algebra of spin $\half N - n$. Geometrically, each set of homogeneous polynomials $\{\varphi_m\}$ corresponding to an irreducible representation defines an algebraic variety $\cV_\varphi$ in $\IC^{N}$ of dimension $2n-1$.

The case of the polynomials $\tilde h_m$ discussed above in~\eqref{hgeneration} simply corresponds to the case $n=2$, where we have $k^2_S=\lambda_S^{(2)}$. For null vectors, all of the coefficients $k_S^2$ can be expressed in terms of the $\half N (N - 1)$ quadric ones $2 k_a \cdot k_b$. In fact, \cite{Dolan:2014ega} showed that spin $\half N - 2$ representations are uniquely defined and, therefore, the polynomials $\varphi_m$ have to have the form of the scattering equation \eqref{scatterhomo} when starting from some highest weight $\varphi_2$. The constraints on the $k$ parameters \eqref{kcons} are precisely those leading to the conditions of highest weight \eqref{HW}. Indeed, combining the null condition $k^2=0$ with the conservation of momenta $\sum k_a=0$, we have,
\be
 \sum_{b\in A \atop b\neq a}k_a\cdot k_b = k_a\cdot\sum_{b\in A} k_b =0\ ,
\ee
leading to,
\be
 \sum_{S\subset A\ |\ a\in S,\atop |S|=2}\lambda_S = 0 \ , \quad \mbox{for each}\quad a\in A \ ,
\ee
with $\lambda_{ab} := 2k_a\cdot k_b$. This is precisely the highest weight conditions \eqref{HW} for $n=2$.

%%%%%%%%%%%%%%%%%%%%%%%%%%%%%%%%%%%%%%%%%

\Mobius invariance allows us to fix some of the $z$ variables of the above homogeneous system of polynomials. Following \cite{Dolan:2014ega}, we choose to set $z_1 \rightarrow \infty$ and $z_N \rightarrow 0$. Fixing these points amounts to acting on every polynomials $\{\varphi_m\}$ with the following operators,

\be\label{fix}
 L_{\rm z_1} = \frac{\partial}{\partial z_1} \ , \quad \mbox{and} \quad L_{z_N}= 1 - z_N \frac{\partial}{\partial z_N} \ ,
\ee
which corresponds to $z_1 \rightarrow \infty$ and $z_N \rightarrow 0$ to respectively. For implementing both conditions, we can act with the product of both operators $L_{\rm z_1}\cdot L_{z_N}$. This has the effect to decrease the degree by $1$ and it is straightforward to realise that the new set of polynomials thus defined is still a subset of the space of \Mobius invariant polynomials. However, they do not lead to another irreducible representation as the highest weight conditions \eqref{HW} are not satisfied for the lowest degree polynomial. We will write the set of polynomials resulting from fixing \Mobius invariance as $\{\varphi^*_m\}$. Explicitly, they are given by,
\be\label{FixDef}
 \varphi^*_m :=\sum_{S\subset A' \atop |S|=m} \lambda^{(n)}_{S^1}z_S \ , \quad \mbox{with} \quad \lambda^{(n)}_{S^1}=\sum_{U \subset {S^1} \atop |U| = n}\lambda_U \ ,
\ee
where $S^1=S\cup \{1\}$ and $A'=\{a\in A : a \neq 1,N\}$. The index range is $n-1\le m\le N-n-1$, therefore, each set contains $N-2n+1$ polynomials. These sets of polynomials $\{\varphi^*_m\}$ thus define algebraic varieties $\cV^*_\varphi$ in $\IC^{N-2}$ of dimension $2n-3$.

The above tensor coefficients $\lambda^{(n)}_{S^1}$ are also subject to the highest weight conditions~\eqref{HW}. However, from the limit $z_N \rightarrow 0$, no tensor $\lambda$ with an index $N$ will appear in~\eqref{FixDef}. Nonetheless, the remaining coefficients cannot be made completely arbitrary. Indeed, they are subject to the constraint,
\be\label{HWfix}
\sum\limits_{U\in A\setminus\{N\}\atop |U|=n}\lambda_U=0 \ ,
\ee
This is the ``highest weight'' condition that hold for any $\cV^*$ varieties. It is obtained considering  \eqref{HW} for $a=N$ and realising that each terms $\lambda_S\ |\ N\in S$ are all individually included in the other constraints coming from $a\ne N$. Thus, the sum of all equations in~\eqref{HW} for $a=1,\dots,N-1$ will contain the vanishing sum of all terms $\lambda_S\ |\ N\in S$. The remaining terms in the sum lead to \eqref{HWfix}.

In the following section, we will aim at describing the geometry of the algebraic varieties defined by the systems of polynomials encountered above. For convenience, let us summarize all the defining equations in Table~\ref{table}.
\begin{table}[h]
\begin{equation*}
\begin{array}{|c|c|c|}\hline
\mbox{Variety} & \mbox{Defining polynomials} & \mbox{Dim (affine)} \\ \hline \hline
\cV_\varphi & 
  \left\{ \varphi_m := \sum\limits_{S\subset A\atop |S|=m} \lambda^{(n)}_S z_S \  | \
  \lambda^{(n)}_S=\sum\limits_{U \subset S \atop |U| = n}\lambda_U \ , \ \sum\limits_{U\subset A\ |\ a\in U,\atop |U|=n  }\lambda_U = 0
  \right\}_{m=n,\ldots,N-n}  & 2n - 1 \\ \hline
\cV^*_\varphi & 
  \left\{ \varphi^*_m := \sum\limits_{S\subset A'\atop |S|=m} \lambda^{(n)}_{S^1}z_S \ | \ \lambda^{(n)}_{S^1}=\sum\limits_{U \subset {S^1}\atop |U| = n}\lambda_U \ , \ \sum\limits_{U\in A\setminus\{N\}\atop |U|=n}\lambda_U=0 \right\}_{m=n-1,\ldots,N-n-1} & 2n - 3 \\ \hline
\end{array}
\end{equation*}
{\caption{ {\sf Summary of defining polynomials for the varieties under consideration. Here, the index set $A = \{1, 2, \ldots, N \}$, $A' = \{a \in A \ | \ a \neq 1, N \}$ and $S^1 = S \cup \{1\}$. The $\lambda$'s are complex coefficients and the varieties are in affine coordinates $z$.}}\label{table}}
\end{table}

Finally, we should note that fixing \Mobius invariance this way decreases the dimension of the variety by two. For the specific case of $n=2$, the (affine) dimension is $1$ and therefore we have a $0$-dimensional (projective) variety, that is, a set of discrete points. The degree of each polynomials is $\{1,\dots,N-3\}$ and from B\'{e}zout's theorem (the number of points corresponds to the product of the degree of the polynomials), we therefore expect for generic choices of the lambda coefficients to have $(N-3)!$ points, counting multiplicity.

%%%%%%%%%%%%%%%%%%%%%%%%%%%%%%%%%%%%%%%%%
%%%%%%%%%%%%%%%%%%%%%%%%%%%%%%%%%%%%%%%%%
\section{Scattering Geometry}\setall

We will now study these algebraic varieties in details in order to grasp their geometrical nature. Noting that for the case $n=2$, $\cV^*_{\varphi}$ is a discrete set of points corresponding to the solutions of the scattering equation, and noting that this is only one specific case defined from irreducible representations of the \Mobius group, we will refer to all the varieties $\cV^*_\varphi$ and $\cV_\varphi$ as {\bf scattering varieties} by abuse of terminology.

Indeed, the deep relation of the scattering equations with irreducible representations of the \Mobius algebra motivates the study of the full class of varieties $\cV_\varphi$ and $\cV_\varphi^*$. However, the question of physical significance of the varieties for $n\ne2$ still remains open. Nevertheless, we hope that an understanding of the geometrical structures will shed light into possible interpretations and the nature of particles scattering. We therefore consider every $\cV_\varphi$ and $\cV_\varphi^*$ without any discrimination.

%%%%%%%%%%%%%%%%%%%%%%%%%%%%%%%%%%%%%%%%%
\subsection{Hilbert series}

Let us first find the dimension, degree and Hilbert series of these varieties.
We start by listing all possibles $N = 2,3,4,5,6,\ldots$ and tabulate the results.
We can extract these geometrical quantities using standard computational geometric packages, such as \cite{sing, mac} as well as software interfacing with Mathematica \cite{Gray:2008zs}.
The tensor coefficients from \eqref{HW} are kept as generic non-vanishing parameters $\lambda_S\ne 0$ satisfying the highest weight condition \eqref{HW}.

The Hilbert series $H(t)$ is a useful tool to identify the nature of an algebraic variety $\cV$. It has a geometrical interpretation in that it supplies a generating function,
\be
H(t) = \sum_{i = 0}^\infty {\rm \dim_{\mathbb{C}}} \, \cV_i\; t^i \ ,
\ee
where the quantity ${\rm \dim_{\mathbb{C}}}\, \cV_i$ is the complex dimension of the graded pieces of $\cV$. It thus represents the number of independent polynomials of degree $i$ on $\cV$ and this encodes information about many geometrical features of the variety. Hilbert series also plays a crucial role in the context of gauge theories~\cite{Benvenuti:2006qr} to count gauge invariant operators.

For convenience of presentation, we will adopt the standard nomenclature that the Hilbert series is presented in the second kind, meaning that for an affine variety $\cV$, we have
\begin{equation}
H(t) = \left({\sum\limits_{i=0}^k a_i t^i}\right) \Big/ {(1-t)^{\dim_{\mathbb{C}}\cV}}  \ ,
\end{equation}
with the power of the denominator being the dimension of the variety and the numerator is a polynomial with integer coefficients $a_i$. (The first kind would have the dimension of the ambient space as the power of the denominator instead).
Furthermore, we will abbreviate the Hilbert series to simply the sequence of coefficients
$
\{a_0,\ a_1,\ \ldots,\ a_k\}
$.
In this second kind, a useful fact is that the sum over $a_i$ (i.e., the numerator evaluated at $t=1$) simply corresponds to the degree of the variety.

%%%%%%%%%%%%%%%
\subsubsection{The variety $\cV_\varphi$}
Using this notation, it is expedient to tabulate our findings of the various Hilbert series obtained through explicit computation \cite{sing, mac}.
First, before imposing the M\"obius transformation, we obtain the results for $\cV_\varphi$ presented in Table \ref{vphi}.
\begin{table}[thb]
\[
{
\begin{array}{|c||c|c|c|}\hline
N & n=2 \, \quad  \dim(\cV_\varphi) = 3 & n=3 \, \quad  \dim(\cV_\varphi) = 5 & 
n=4 \, \quad  \dim(\cV_\varphi) = 7 \\ \hline \hline
4 & 1,1 & - & - \\ \hline
5 & 1,2,2,1 & - & -\\ \hline
6 & 1,3,5,6,5,3,1 & 1,1,1 & - \\ \hline
7 & 1,4,9,15,20,22,20,15,9,4,1 & 1,2,3,3,2,1 & - \\ \hline
8 & \ldots & 1,3,6,9,11,11,9,6,3,1 & 1,1,1,1 \\ \hline
9 & \ldots & \ldots & 1,2,3,4,4,3,2,1  \\ \hline
10 & \ldots & \ldots & 1,3,6,10,14,17,18,17,14,10,6,3,1\\
\hline
\end{array}
}
\]
{\caption{\label{vphi} {\sf Dimension and Hilbert series for $\cV_\varphi$.
Note that we have recorded the affine dimension here, whereby embedding $\cV_\varphi$ into $\IC^{N}$.}}}
\end{table}

A few observations are immediate. First, from combinatorics, all the numerators are {\bf palindromic} in that $a_i = a_{k-i}$ for all $i$. This means, by a theorem of Stanley \cite{stanley,Forcella:2008bb}, that all the corresponding varieties $\cV_\varphi$ are, in fact, {\it affine Calabi-Yau}.\footnote{Strictly speaking, the module generated by the ideal should be a Cohen-Macaulay graded integral domain for this theorem to hold. This will always the case for the varieties considered and we therefore identify palindromic Hilbert series with Calabi--Yau geometries, where Calabi--Yau is meant in the sense of a trivial canonical sheaf. We refer the reader to, for instance, the appendices of~\cite{He:2014oha} for more details on such considerations.}

Next, the sequences of numbers for $n=2$ and $n=3$ are well-known.
The $n=2$ case corresponds to the so-called {\it Mahonian triangle}, the triangle of Mahonian numbers $T_{p,k}$. One combinatorial definition of these numbers \cite{oeis} is that it is the number of permutations $\pi = (\pi(1),\ldots,\pi(p))$ of $\{1,\ldots,p\}$ such that the so-called major index $\sum\limits_{\pi(i)>\pi(i+1)} i$ is equal to $k$. They have a nice generating function which allows us to analytically write the Hilbert series as a function of $N \ge 4$:
\begin{equation}
H(t; N)_{n=2} = (1-t)^{-3} \prod\limits_{j=1}^{N-3} \sum\limits_{i=0}^j t^i = (1-t)^{-3}\sum_kT_{N-3,k}t^k\ ,
\end{equation}
This nice analytical formula deduced from our examples allows us to conjecture its validity for any number of scattering particles $N$.

Similarly, the $n=3$ case corresponds to a $k$-generalization of lattice permutations. As a function of $N \ge 7$ (the initial case of $N=6$ has the numerator $1+t+t^2$ which does not obey the following generating function), we have:
\begin{equation}
H(t; N)_{n=3} = (1-t)^{-5} \prod\limits_{j=1}^{N-4} C_{j+1}(t) \ ; \qquad
C_{j}(t) := \prod\limits_{0<k<j, \ \gcd(k,j)=1} ( t - e^{2 \pi i\frac{k}{j}}) \ ,
\end{equation}
where, in the above, $C_{j}(t)$ are the {\it cyclotomic polynomials}. Again, a nice analytical formula supports the conjecture for its validity up to any $N$ value.\footnote{It would be interesting to find similar constructions for the following cases with $n \geq 4$, which is still eluding the authors.}

%%%%%%%%%%%%%%%
\subsubsection{The variety $\cV^*_\varphi$}
After fixing variables of the above varieties by two M\"obius transformations, $z_1 \to \infty$ and $z_N \to 0$, the complex (affine) dimension of the variety uniformly drops by 2 and we obtain the algebraic varieties $\cV^*_\varphi$. Again, Hilbert series can be obtained through explicit computation~\cite{sing, mac} and the resulting Hilbert series are summarized in Table \ref{vphifix}.
\begin{table}[ht!!]
\[
{
\begin{array}{|c||c|c|c|}\hline
N & n=2 \, \quad  \dim(\cV^*_\varphi) = 1 & n=3 \, \quad  \dim(\cV^*_\varphi) = 3 & 
n=4 \, \quad  \dim(\cV^*_\varphi) = 5 \\ \hline \hline
4 & 1 & - & - \\ \hline
5 & 1,1 & - & - \\ \hline
6 & 1,2,2,1 & 1,1 & -\\ \hline
7 & 1,3,5,6,5,3,1 & 1,2,2,1 & - \\ \hline
8 & 1,4,9,15,20,22,20,15,9,4,1 & 1,3,5,6,5,3,1 & 1,1,1 \\ \hline
9 & \ldots & 1,4,9,15,20,22,20,15,9,4,1 & 1,2,3,3,2,1 \\ \hline
10 & \ldots & \ldots & 1,3,6,9,11,11,9,6,3,1  \\ \hline
\end{array}
}
\]
{\caption{\label{vphifix} {\sf Affine dimension and Hilbert series for $\cV^*_\varphi$}.}}
\end{table}

We observe that these Hilbert series are simply a shift of the above table. For the case $n=2$, we recover the same Hilbert series as for the $\cV_\varphi$ case, shifted from $N$ to $N+1$. For the other cases, the Hilbert series for $n$ correspond to that of $\cV_\varphi$ for $n-1$ shifted from $N$ to $N+2$. This is reminiscent from the fact that the number of variables decreases by $2$ when \Mobius invariance is fixed. Thus, we find again the Mahonian numbers, however now for both the $n=2$ and $n=3$ cases, and the cyclotomic polynomials for the $n=4$ case.

This observation unveils deep connections across the varieties $\cV_\varphi$ and $\cV^*_\varphi$ for different particle numbers $N$. We should note, however, that Hilbert series are not topological invariants as it can be presented in many ways. In fact, they depend on the embedding of the variety within the polynomial ring. This means that we cannot conclude that we have identical varieties when they share the same Hilbert series. Nevertheless, the Calabi--Yau property is deducible from the palindromic nature and it is remarkable that that the scattering varieties are all affine Calabi--Yau manifolds without recourse to supersymmetry or string theory.

%%%%%%%%%%%%%%%
\subsection{Resolution and Betti numbers}

Having established the correspondence between the dimension of the graded pieces of $\cV_\varphi$ and $\cV_\varphi^*$ via the Hilbert series, let us see whether we can highlight further similarities or differences.
For instance, one could examine the resolution of the module $M = R/V$ where $R$ the polynomial ring $\IC[z_1,\dots,z_N]$ for $\cV_\varphi$ and $\IC[z_2,\dots,z_{N-1}]$ for $\cV_\varphi^*$ and $V$ is the corresponding variety ideal. Remarkably, it turns out that
the Betti numbers corresponding to this resolution follow a predictable pattern as we now see. We emphasize that the Betti number here are not the usual topological Betti numbers of which one is more familiar. The Betti numbers henceforth refers to the notation in algebraic geometry of the dimensions of the modules in a free resolution of a given ideal\footnote{More information on the meaning of Betti numbers of minimal resolution of zero-dimensional ideals can be found in~\cite{anna}.}.

Let us start by exemplifying the resolution of the module corresponding to the somewhat trivial case of $\cV_\varphi^*$ for $n=2$, e.g. the discrete set of points. Considering $N=4$, we have the following ideal:
\begin{equation}
V^*_4 = \langle \lambda_{12}z_2+\lambda_{13}z_3 \rangle \ ,
\end{equation}
with the tensor coefficients $\lambda_{12}$ and $\lambda_{13}$ subject to the highest weight condition~\eqref{HWfix}. In this case, this translates to $\lambda_{12}+\lambda_{13}+\lambda_{23}=0$, and, since $\lambda_{23}$ does not appear in the ideal, the parameters $\lambda_{12}$ and $\lambda_{13}$ can be considered as generic parameters (provided that $\lambda_{23}$ takes the adequate value).
The minimal resolution of the module $M^*_4 = \mathbb{C}[z_2,z_3]/V_4^*$ gives to the short exact sequence
\begin{equation}
0\xleftarrow[]{{\quad}}M^*_4\xleftarrow[\quad]{{}}{R^1}\xleftarrow[z_2+(\lambda_{13}/\lambda_{12})z_3]{}{R^1}\xleftarrow[]{\quad}{0} \,
\end{equation}
with the Betti numbers
$$
\begin{array}{c||c|c}
i \backslash j & 0 & 1  \\ \hline \hline
0 & 1&1 \\ \hline
%1 & 0&1 \\ \hline\hline
{\rm Total} & 1 & 1
\end{array}
$$
This tally means that the $j$-th column of the $i$-th row gives the number of basis elements of degree $i+j$ in the free module $M^*_4[j]$ of shifted degree $j$ (e.g., in this case, $M^*_4[0]=R^1$ and $M^*_4[1]=R^1$). The total corresponds to the Betti numbers, giving the number of copies of the polynomial ring $R = \mathbb{C}[z_2,z_3]$, meaning $R^i = \oplus_i R$ with $i$ the corresponding Betti number.

Using~\cite{sing, mac}, we easily obtain the tally for the Betti numbers of resolutions of other modules $M_N^*$. For instance, still considering $n=2$, we find for $N=5$ and $N=6$, respectively:

\begin{align*}
&N=5 &&N=6\\
&\begin{array}{c||c|c|c}
i \backslash j & 0 & 1 & 2 \\ \hline \hline
0 & 1&1&0 \\ \hline
1 & 0&1&1 \\ \hline\hline
{\rm Total} &1&2&1
\end{array}
%\hspace{1cm}
&&\begin{array}{c||c|c|c|c}
i \backslash j & 0 & 1 & 2 & 3\\ \hline \hline
0&  1&1&0&0 \\ \hline
1&  0&1&1&0 \\ \hline
2&  0&1&1&0 \\ \hline
3&  0&0&1&1 \\ \hline\hline
{\rm Total} &1&3&3&1
\end{array}
\end{align*}

What about the $\cV$ varieties? For comparison, let us present the tally corresponding to the Betti numbers of the resolution of the module defining the varieties $\cV_\varphi$ for $n=2$ and $N=4$, $N=5$ and $N=6$. We obtain:

\begin{align*}
&N=4 &&N=5 &&N=6\\
&\begin{array}{c||c|c}
i \backslash j & 0 & 1  \\ \hline \hline
0 & 1&0 \\ \hline
1 & 0&1 \\ \hline\hline
{\rm Total} & 1 & 1
\end{array}
&&\begin{array}{c||c|c|c}
i \backslash j & 0 & 1 & 2 \\ \hline \hline
0 & 1&0&0 \\ \hline
1 & 0&1&0 \\ \hline
2 & 0&1&0 \\ \hline
3 & 0&0&1 \\ \hline\hline
{\rm Total} &1&2&1
\end{array}
%\hspace{1cm}
&&\begin{array}{c||c|c|c|c}
i \backslash j & 0 & 1 & 2 & 3\\ \hline \hline
0&  1&0&0&0 \\ \hline
1&  0&1&0&0 \\ \hline
2&  0&1&0&0 \\ \hline
3&  0&1&1&0 \\ \hline
4&  0&0&1&0 \\ \hline
5&  0&0&1&0 \\ \hline
6&  0&0&0&1 \\ \hline\hline
{\rm Total} &1&3&3&1
\end{array}
\end{align*}

Interestingly, we notice that the total matches the one for the varieties where \Mobius invariance has been fixed, albeit the tally being different. In fact, this is quite general. We summarise the Betti numbers in Table~\ref{Betti}. Both the varieties $\cV_\varphi$ and $\cV_\varphi^*$ share the same Betti numbers for the resolution of their module, when they have equal number of particles $N$. Moreover, the sequence is repeated for various $n$, with a shift $N \rightarrow N + 2$ when $n \rightarrow n + 1$.

\begin{table}[h]
\[
{
\begin{array}{|c|c|c||c|}\hline
\multicolumn{3}{|c||}{N}  & \multirow{2}{*}{\mbox{Betti numbers}} \\
\cline{1-3}
n=2&  n=3 & n=4 & \\ \hline \hline
4 & 6 & 8 & 1,1 \\ \hline
5 & 7 & 9 & 1,2,1 \\ \hline
6 & 8 & 10 & 1,3,3,1 \\ \hline
7 & 9 & \ldots & 1,4,6,4,1 \\ \hline
8 & \ldots &  & 1,5,10,10,5,1 \\ \hline
\ldots &  &  & \ldots \\ \hline
\end{array}
}
\]
{\caption{\label{Betti} {\sf Betti numbers of the module free resolution for the 
varieties $\cV_\varphi$ and $\cV_\varphi^*$, with respect to the number of particles $N$ and the highest weight degree $n$.}}}
\end{table}

We clearly see a pattern and we can conjecture that it holds for every remaining varieties. The minimal resolution of the modules $M$ and $M^*$ will be an exact sequence with the numbers of generators forming Pascal's triangle; a generating function for the Betti numbers is then obtained from $(1+x)^{N-2n+1}$.

With these Betti numbers, we can also note that the length of the resolution is equal to $N-3$. Thus, we always have the codimension of $M$ (and $M^*$) equal to the length of its resolution, thus all the quotient rings considered are arithmetically Cohen--Macaulay.

%%%%%%%%%%%%%%%%%%%%%%%%%%%%%%%%%%%%%%%%%
\subsection{Calabi--Yau geometry}

Are the varieties of geometrical significance, or, at least, of some familiarity?
Let us focus on the cases of affine dimension 3. These are the cases with $n=2$ and $n=3$, respectively before and after \Mobius fixing, with a shift of two in $N$.

The first variety encountered has $H(t) = (1-t)^{-3}(1+t)$. Guided by the plethystic logarithm \cite{Benvenuti:2006qr}, let us recast this into the Euler form, viz., the form with both the numerator and the denominator is factorized as a product of $(1-t^i)$ for some $i$ coefficients.  We obtain $H(t) = (1-t)^{-4}(1-t^2)$ and can readily identify \cite{Benvenuti:2006qr} the Hilbert series of the famous {\it conifold} as as an affine (non-compact) CY3 , given by a quadric in $\IC^4$.

The next variety has $H(t) = (1-t)^{-3}(1 + 2t + 2t + t^2)$ which is, when put into the Euler form, $H(t) = (1-t^2)(1-t^3) / (1-t)^5$. This means we have the intersection of a quadric and a cubic in $\IC^5$. Upon projectivizing to $\IC\IP^4$, this is a complete intersection which gives a K3 surface of degree 6 and (geometric) genus 4 \cite{brown}.

Continuing in a similar fashion, we can identify all the dimension 3 (projective dimension 2) cases as complete intersection complex surfaces
and use the standard notation \cite{hubsch}:
\begin{equation}
[n | k_1, k_2, \ldots, k_m] := \{\mbox{intersection of $m$ degree $k_i$ hypersurfaces in $\IC\IP^n$}\}  \ .
\end{equation}
With this notation, the K3 condition, that is vanishing of the first Chern class, is equivalent to $\sum_{i=1}^m k_i = n + 1$. Similarly, a condition for being a Fano variety, is that the left-hand side of this expression is greater, that is $\sum_{i=1}^m k_i > n + 1$.
We can thus summarize the (projective) geometries as in Table \ref{geom}.
\begin{table}[thb]
\[
\begin{array}{|c|c|c|c|}\hline
\mbox{Geometry} & \mbox{Type} & \mbox{Hilbert Series} & \mbox{Degree}  \\
\hline \hline
[3|2] & \mbox{Base of conifold} & 1,1 & 2!  \\ \hline
[4|2,3] & \mbox{K3} & 1,2,2,1 & 3!  \\ \hline
[5|2,3,4] & \mbox{Fano} &  1,3,5,6,5,3,1 & 4! \\ \hline
[6|2,3,4,5] & \mbox{Fano} & 1,4,9,15,20,22,20,15,9,4,1 & 5!  \\ \hline
\end{array}
\]
{\caption{\label{geom} {\sf Geometry of the affine $3$-dimensional (and hence, projective dimension 2) varieties. The {\it Type} refers to the base $\cB$ after projectivisation of the varieties}.}}
\end{table}

The regularity of this table easily allows to speculate on the nature of the following geometries for greater $N$. We expect thus all remaining $3$-dimensional algebraic varieties to be Fano, as presented in Table \ref{geomspec}.
\begin{table}[h]
\[
\begin{array}{|c|c|c|c|c|}\hline
&\mbox{Geometry} & \mbox{Type} & \mbox{Hilbert Series} & \mbox{Degree}  \\
\hline \hline
\cV_\varphi &[N-1|2,3,4,\dots,N-2] & \mbox{Fano} &  \left\{T_{N-3,k}\right\}_{\forall k} & (N-2)! \\ \hline
\cV_\varphi^* &[N-3|2,3,4,\dots,N-4] & \mbox{Fano} & \left\{T_{N-5,k}\right\}_{\forall k} & (N-4)!  \\ \hline
\end{array}
\]
{\caption{\label{geomspec} {\sf General description of the affine $3$-dimensional geometry for $\cV_\varphi$ with $N\geq 6$ and $\cV_\varphi^*$ with $N\geq 8$}.}}
\end{table}

Moreover, we should emphasize that, by the palindromic nature of the Hilbert series in the  second kind, the varieties are all affine Calabi-Yau threefolds as affine (unprojectivized) $3$-dimensional varieties. 
They can therefore be considered as complex cones over some compact base surface, precisely in the same way as in the Calabi-Yau singularities of $AdS_5/CFT_4$.

%%%%%%%%%%%%%%%%%%%%%%%%%%%%%%%%%%
\subsection{Singular locus}

One might wonder then, what is the difference between the varieties $\cV_\varphi$ and $\cV_\varphi^*$ from a geometrical standpoint? Calculating the Euler number $\chi$ of the above projective varieties unveils more information about the nature of the geometry considered. In this section, we concentrate our study on the cases of $3$-dimensional varieties, that is $\cV_\varphi$ for $n=2$ and $\cV_\varphi^*$ for $n=3$. Our findings for $\chi$ using~\cite{sing, mac} are presented in Table \ref{Eul}.
\begin{table}[h]
\[
\begin{array}{|c|c||c|c||c|}\hline
\multicolumn{2}{|c||}{\mbox{$\cV_\varphi$ with $n=2$}}&\multicolumn{2}{c||}{\mbox{$\cV_\varphi^*$ with $n=3$}}&\multirow{2}{*}{\mbox{Type}}\\  \cline{1-4}
 \mbox{$N$} & \chi &\mbox{$N$} & \chi &\\ \hline \hline
 4&4&6&4&\mbox{Base of conifold}\\ \hline
 5&29&7&24&\mbox{K3}\\ \hline
 6&502&8&384&\mbox{Fano}\\ \hline
\end{array}
\]
{\caption{\label{Eul} {\sf Euler numbers $\chi$ for the $3$-dimensional varieties encountered}.}}
\end{table}

In fact, we can go further and compute the full Hodge diamonds for these varieties. We find,
\begin{equation}
h^{p,q}(\cB_{4, 2}) \quad = \quad
{\begin{array}{ccccc}
&&h^{0,0}&& \\
&h^{0,1}&&h^{0,1}& \\
h^{0,2}&&h^{1,1}&&h^{0,2} \\
&h^{0,1}&&h^{0,1}& \\
&&h^{0,0}&& \\
\end{array}}
\quad = \quad
{\begin{array}{ccccc}
&&1&& \\
&0&&0& \\
0&&2&&0 \\
&0&&0& \\
&&1&& \\
\end{array}} \ ,
\label{hodge1}
\end{equation}
where we use the notation $\cB_{N, n}$ for the projective variety corresponding to $\cV_\varphi$ with $N$ particles and highest weight polynomial of degree $n$. This Hodge diamond corresponds to the one of the conifold, as expected. Furthermore, we find that
\be
h^{p,q}(\cB^*_{6, 3})=h^{p,q}(\cB_{4, 2})
\ee
for the corresponding variety $\cB^*$ resulting from fixing \Mobius invariance. The Hodge diamonds for the other varieties present in Table~\ref{Eul} are, for the K3 surfaces:
\begin{equation}
h^{p,q}(\cB_{5, 2})  = 
{\begin{array}{ccccc}
&&1&& \\
&0&&0& \\
1&&25&&1 \\
&0&&0& \\
&&1&& \\
\end{array}} \ , \qquad  
h^{p,q}(\cB^*_{7, 3})  = 
{\begin{array}{ccccc}
&&1&& \\
&0&&0& \\
1&&20&&1 \\
&0&&0& \\
&&1&& \\
\end{array}} \ ,
\label{hodge1}
\end{equation}
and for the Fano varieties:
\begin{equation}
h^{p,q}(\cB_{6, 2})  = 
{\begin{array}{ccccc}
&&1&& \\
&0&&0& \\
49&&402&&49 \\
&0&&0& \\
&&1&& \\
\end{array}} \ , \qquad  
h^{p,q}(\cB^*_{8, 3})  = 
{\begin{array}{ccccc}
&&1&& \\
&0&&0& \\
49&&284&&49 \\
&0&&0& \\
&&1&& \\
\end{array}} \ .
\label{hodge1}
\end{equation}

The astute reader would see the Hodge diamond and the Euler characteristic $29$ corresponding to the K3 surface and be puzzled.
Here, we have a K3 surface as a complete intersection of a quadric and a cubic in $\IC\IP^4$. Should the Euler number then not be the standard 24?
Upon inspection of the defining polynomials from \eqref{poly}, we see that the tensor coefficients $\lambda$ to these polynomials in $z$ are not generic and obey the physical constraints~\eqref{HW}. In other words, we are not in a generic point in the complex structure moduli space of K3 surfaces but quite special ones. In fact, we will now see that the scattering varieties $\cV_\varphi$ for $n=2$ all contain singularities and are therefore not smooth, which will deviate the values of the Euler number from generic varieties.

The singular locus of an algebraic variety corresponds to its intersection with the Jacobian ideal, i.e., the zero locus of its Jacobian matrix, giving the variety, for that defined by \eqref{poly},
\be
J = \left( {\begin{array}{*{20}{c}}
  {\frac{{\partial {\varphi_n}}}{{\partial {z_1}}}}&{\ldots}&{\frac{{\partial {\varphi_n}}}{{\partial {z_N}}}} \\ 
  {\vdots}&{\ldots}&{\vdots} \\ 
  {\frac{{\partial {\varphi_{N - n}}}}{{\partial {z_1}}}}&{\ldots}&{\frac{{\partial {\varphi_{N - n}}}}{{\partial {z_N}}}} 
\end{array}} \right) = 0 \ .
\ee
Let us write indices $i = 1,\ldots,N$ for $z$ and $m=n,\ldots,N-n$ for $\varphi$. Therefore, we have 
\be\label{Jaco}
\frac{{\partial \varphi_m}}{{\partial {z_i}}} = \sum\limits_{S\subset A\ |\ i\in S,\atop |S|=m} {\lambda _S^{(n)}{z_{S\setminus \{i\} }}} \ .
\ee
We should realise that each polynomial from the Jacobian corresponds to the limit where the variable $z_i \rightarrow \infty$ as this is the effect of the action from the operators $L_{\rm z_i} = \frac{\partial}{\partial z_i}$, as can be seen by generalisation of the $z_1$ variable case~\eqref{fix}.
From the ideal generated by these partial derivatives --- the Jacobian ideal --- we therefore have $N-2n+1$ polynomials of degree ranging from $n-1$ to $N-n-1$, for each $i=1,\ldots,N$, defining an algebraic variety in $\IC\IP^{N-1}$.

Let us focus on the important case of $n=2$.
First, for $N=4$, we have only one homogeneous degree 2 polynomial $\{ {\tilde h_2} \}$. Explicitly,
\be
{V_4}  =    \langle {\lambda _{12}}{z_1}{z_2} + {\lambda _{13}}{z_1}{z_3} + {\lambda _{14}}{z_1}{z_4} + {\lambda _{23}}{z_2}{z_3} + {\lambda _{24}}{z_2}{z_4} + {\lambda _{34}}{z_3}{z_4}\rangle \ .
\ee
The corresponding Jacobian ideal is then
\begin{align}\nn
{J_4} = \langle &
  {{\lambda _{12}}{z_2} + {\lambda _{13}}{z_3} + {\lambda _{14}}{z_4}} ,
  {{\lambda _{12}}{z_1} + {\lambda _{23}}{z_3} + {\lambda _{24}}{z_4}} , \\
&
  {{\lambda _{13}}{z_1} + {\lambda _{23}}{z_2} + {\lambda _{34}}{z_4}} , 
  {{\lambda _{14}}{z_1} + {\lambda _{24}}{z_2} + {\lambda _{34}}{z_3}} 
\rangle \ .
\end{align}
We readily see that for arbitrary generic choices of $\lambda$, the only point in $J_4$ is when all $z_i = 0$, which in $\IC^4$ is the origin and in $\IC\IP^3$ is excluded.
Hence, for this generic case, the quadric projective variety $[3|2]$ is smooth and as an affine variety is realized as the conifold, with the familiar singularity at the origin corresponding to the tip of the cone.

However, for irreducible representation of the \Mobius group, the tensor coefficients $\lambda$ are not generic. They must satisfy the constraints of highest weight~\eqref{HW}, that is,
\be\label{V4HW}
\left\{
\begin{gathered} {\lambda _{12}} + {\lambda _{13}} + {\lambda _{14}} = 0 \hfill \; ,\\ {\lambda _{12}} + {\lambda _{23}} + {\lambda _{24}} = 0 \hfill \; , \\ {\lambda _{13}} + {\lambda _{23}} + {\lambda _{34}} = 0 \hfill \; ,\\ {\lambda _{14}} + {\lambda _{24}} + {\lambda _{34}} = 0 \hfill \; .\\ \end{gathered}\right.
\ee
In this case, we can find a one-parameter family of solutions to $V_4$ and $J_4$ given by the following,
\begin{equation}
(z_1,z_2,z_3,z_4) = (x,x,x,x) \ , \quad x \in \IC \ .
\end{equation}
This gives the {\bf singular locus} on our non-generic variety.
On the affine variety, the singular locus is a ray from the origin while, on the projective variety, it is a single point $[1:1:1]$ on the quadric in $\IC\IP^3$.

Next, for the $N=5$ case, we have a quadric intersecting a cubic. Explicitly,
\begin{eqnarray}\nn
{V_5} & = & \langle \ 
{\lambda _{12}}{z_1}{z_2} + {\lambda _{13}}{z_1}{z_3} + {\lambda _{14}}{z_1}{z_4} + {\lambda _{15}}{z_1}{z_5} + {\lambda _{23}}{z_2}{z_3}  \\ \nn
&& + {\lambda _{24}}{z_2}{z_4} + {\lambda _{25}}{z_2}{z_5} + {\lambda _{34}}{z_3}{z_4} + {\lambda _{35}}{z_3}{z_5} + {\lambda _{45}}{z_4}{z_5} \ ,  \\ \nn
&& (\lambda _{12}+\lambda _{13}+\lambda _{23}){z_1}{z_2}{z_3} + (\lambda _{12}+\lambda _{14}+\lambda _{24}){z_1}{z_2}{z_4} \\ \nn
&& + (\lambda _{12}+\lambda _{15}+\lambda _{25}){z_1}{z_2}{z_5} + (\lambda _{13}+\lambda _{14}+\lambda _{34}){z_1}{z_3}{z_4}  \\  \nn
&& + (\lambda _{13}+\lambda _{15}+\lambda _{35}){z_1}{z_3}{z_5} + (\lambda _{14}+\lambda _{15}+\lambda _{45}){z_1}{z_4}{z_5}  \\ \nn
&& + (\lambda _{23}+\lambda _{24}+\lambda _{34}){z_2}{z_3}{z_4} + (\lambda _{23}+\lambda _{25}+\lambda _{35}){z_2}{z_3}{z_5}   \\
&& + (\lambda _{24}+\lambda _{25}+\lambda _{45}){z_2}{z_4}{z_5} + (\lambda _{34}+\lambda _{35}+\lambda _{45}){z_3}{z_4}{z_5}   \
\rangle
\ ,
\end{eqnarray}
where the highest weight~\eqref{HW} constraints on the coefficients are:
\begin{equation}\label{HWforV5}
\left\{
\begin{gathered}
{\lambda _{12}} + {\lambda _{13}} + {\lambda _{14}} + {\lambda _{15}} = 0 \hfill \ , \\ 
{\lambda _{12}} + {\lambda _{23}} + {\lambda _{24}} + {\lambda _{25}} = 0 \hfill \ , \\ 
{\lambda _{13}} + {\lambda _{23}} + {\lambda _{34}} + {\lambda _{35}} = 0 \hfill \ , \\ 
{\lambda _{14}} + {\lambda _{24}} + {\lambda _{34}} + {\lambda _{45}} = 0 \hfill \ , \\
{\lambda _{15}} + {\lambda _{25}} + {\lambda _{35}} + {\lambda _{45}} = 0 \hfill \ . \\
\end{gathered} \right. 
\end{equation}
This gives a non-generic K3 surface upon projectivization. The corresponding Jacobian ideal is given by

\begin{eqnarray} \nn
J_5 & = & \langle \ \lambda _{12} z_2+\lambda _{13} z_3+\lambda _{14} z_4+\lambda _{15} z_5,  \ \lambda _{12}z_1 +\lambda _{23} z_3+\lambda _{24} z_4+\lambda _{25} z_5, \\
\nn
 &&    \lambda _{13}z_1+ \lambda _{23}z_2+\lambda _{34} z_4+\lambda _{35} z_5,  \ \lambda _{14}z_1+ \lambda _{24}z_2+ \lambda _{34}z_3+\lambda _{45} z_5, \\
\nn
 &&    \lambda _{15}z_1+ \lambda _{25}z_2+ \lambda _{35}z_3+ \lambda_{45}z_4,  \ (\lambda _{12}+\lambda _{13}+\lambda _{23}) z_2 z_3 \\
\nn
 &&  + (\lambda _{12}+\lambda _{14}+\lambda _{24}) z_2 z_4 +(\lambda _{12}+\lambda _{15}+\lambda _{25}) z_2 z_5 + (\lambda _{13}+\lambda _{14}+\lambda _{34}) z_3 z_4 \\
\nn
 &&+ (\lambda _{13}+\lambda _{15}+\lambda _{35}) z_3 z_5 +(\lambda _{14}+\lambda _{15}+\lambda _{45}) z_4 z_5,  \  (\lambda _{12}+\lambda _{13}+\lambda _{23})z_1 z_3\\
\nn
 && + (\lambda _{12}+\lambda _{14}+\lambda _{24})z_1 z_4 + (\lambda _{12}+\lambda _{15}+\lambda _{25})z_1z_5 +(\lambda _{23}+\lambda _{24}+\lambda _{34}) z_3 z_4\\
\nn
 &&+(\lambda _{23}+\lambda _{25}+\lambda _{35}) z_3 z_5  +(\lambda _{24}+\lambda _{25}+\lambda _{45}) z_4 z_5, \ (\lambda _{12}+\lambda _{13}+\lambda _{23})z_1 z_2\\
\nn
 &&+ (\lambda _{13}+\lambda _{14}+\lambda _{34}) z_1z_4  + (\lambda _{13}+\lambda _{15}+\lambda _{35}) z_1z_5 + (\lambda _{23}+\lambda _{24}+\lambda _{34}) z_2z_4\\
\nn
 &&+ (\lambda _{23}+\lambda _{25}+\lambda _{35}) z_2z_5 +(\lambda _{34}+\lambda _{35}+\lambda _{45}) z_4 z_5, \   (\lambda _{12}+\lambda _{14}+\lambda _{24}) z_1z_2\\
\nn
 &&+ (\lambda _{13}+\lambda _{14}+\lambda _{34})z_1 z_3 + (\lambda _{14}+\lambda _{15}+\lambda _{45}) z_1z_5 + (\lambda _{23}+\lambda _{24}+\lambda _{34}) z_2z_3\\
\nn
 &&+ (\lambda _{24}+\lambda _{25}+\lambda _{45})z_2 z_5+ (\lambda _{34}+\lambda _{35}+\lambda _{45}) z_3 z_5,  \  (\lambda _{12}+\lambda _{15}+\lambda _{25}) z_1z_2\\
\nn
 &&+ (\lambda _{13}+\lambda _{15}+\lambda _{35})z_1 z_3+ (\lambda _{14}+\lambda _{15}+\lambda _{45}) z_1z_4 + (\lambda _{23}+\lambda _{25}+\lambda _{35})z_2 z_3\\ \label{J5}
&&+ (\lambda _{24}+\lambda _{25}+\lambda _{45})z_2 z_4+ (\lambda _{34}+\lambda _{35}+\lambda _{45}) z_3z_4 \ \rangle \ .
\end{eqnarray}
Again, looking at the terms in $J_5$ linear in $z$, we see that on the ray from the origin,
\be
(z_1,z_2,z_3,z_4,z_5) = (x,x,x,x,x) , \quad x \in \IC \, ,
\ee
the solution set of the highest weight conditions \eqref{HWforV5} is exactly what is required to make the five linear polynomials in~\eqref{J5} vanish.
In fact, one can see that for all $N$, the set of linear constraints on the coefficients $\lambda$ will exactly make the Jacobian vanish for the ray $(x,x,\ldots,x)$ whereby making the projective point $[1:1:\ldots:1]$ always a singular point on the scattering varieties $\cV_\varphi$ for $n=2$.

To demonstrate this, let us write the highest weight \eqref{HW} condition for $n=2$ as follows:
\begin{equation}\label{hwcn2}
\left\{\sum\limits_{i \in A \backslash  \{ r\} } {{\lambda _{ri}}}  = 0\right\}_{\forall r \in A}
\end{equation}
This constitutes of a set of $N$ equations for each $r$ in $A$.
Now, for one specific constraint with index $r'$, each term $\lambda_{r'i}$ with $i\in A\backslash \{r'\}$ will appear once (and only once) in the remaining set of constraints (with $r\neq r'$). Indeed, this is the case when $r$ hits the corresponding $i$ index. This means that summing all equations for which $r\neq r'$ and removing the one with $r=r'$ leads to a constraint involving only the terms $\lambda_{ij}$ for which $i,j \in A\backslash \{r'\}$. (Each of those terms will appear twice from symmetry of the indices.) 
This gives another set of identities:
\begin{equation}\label{hwcn3}
\left\{\sum\limits_{i,j \in A \backslash \{ r\} ,\; i < j} {{\lambda _{ij}}}  = 0\right\}_{\forall r \in A}
\end{equation}
As an illustration, we can take the case $N=5$ for which the highest weight conditions are explicitly given in \eqref{HWforV5}. Let us choose the index $r'=1$. Adding the last four equations from \eqref{HWforV5} minus the first one leads to $2({\lambda _{23}} + {\lambda _{24}} + {\lambda _{25}} + {\lambda _{34}} + {\lambda _{35}} + {\lambda _{45}} )= 0$ and we see that the sum of all $\lambda$ terms with indices in $A \backslash \{1\}$ must vanish, as stated in \eqref{hwcn3}. The remaining equations in \eqref{hwcn3} follows from the different choices of $r'$.

We should also recast the Jacobian ideal in the following form (altogether $N(N-3)$ polynomials):
\begin{equation}\label{Jideal}
\begin{array}{*{20}{c}}
  {\sum\limits_{p \in A \backslash \{ r\} } {(\sum\limits_{i < j|\{ i,j\}  \subset \{ p,r\} } {{\lambda _{ij}}} } )\; {z_p} = 0, \quad \forall r \in A }\ , \\ 
  {\sum\limits_{p,q \in A \backslash \{ r\} } {(\sum\limits_{i < j|\{ i,j\}  \subset \{ p,q,r\} } {{\lambda _{ij}}} } )\; {z_p}{z_q} = 0,\quad \forall r \in A }\ ,\\ 
  {\sum\limits_{p,q,u \in A \backslash \{ r\} } {(\sum\limits_{i < j|\{ i,j\}  \subset \{ p,q,u,r\} } {{\lambda _{ij}}} } )\; {z_p}{z_q}{z_u} = 0,\quad \forall r \in A} \ ,\\ 
  {\vdots} \\ 
  {\sum\limits_{{p_1},...,{p_{N - 3}} \in A \backslash \{ r\} } {(\sum\limits_{i < j|\{ i,j\}  \subset \{ {p_1},...,{p_{N - 3}},r\} } {{\lambda _{ij}}} } )\; {z_{{p_1}}}...{z_{{p_{N - 3}}}} = 0,\quad \forall r \in A}\ .
\end{array}
\end{equation}

Our claim is that, after taking all the $z$ variables to be $x \in \IC$, each condition in (\ref{Jideal}) are automatically satisfied. Indeed, in this case, we can factor out all the $x$ variables and the first $N$ equations with degree $1$ simply vanish by virtue of the highest weight conditions~\eqref{HW}.
For the remaining equations of degree $d > 1$, the conditions can be rewritten as follows after some combinatorial reorganisation:
\be\left(\binom{N - 2}{d - 1}\sum\limits_{i \in A \backslash \{ r\} } {{\lambda _{ir}}}  + \binom{N - 3}{d - 2}\sum\limits_{p,q \in A \backslash \{ r\},\;  p < q} {{\lambda _{pq}}} \right)\; {x^d}= 0\; ,\quad \forall r \in A \ .
\ee
We can see that the coefficients involving the $\lambda$ terms vanish since we have $\sum\limits_{i \in A \backslash \{ r\} } {{\lambda _{ir}}} = 0 $ from \eqref{hwcn2} and $\sum\limits_{p,q \in A \backslash \{ r\},\;  p < q} {{\lambda _{pq}}} =0$ from \eqref{hwcn3}. Therefore all the equations \eqref{Jideal} are satisfied for the point $(x,x,\ldots,x)$, making the projective point $[1:1:\ldots:1]$ always singular.

It is remarkable that the existence of this singularity is deeply rooted into the constraint from highest weight \eqref{HW}, and hence the fact that the polynomial system form an irreducible representation of the \Mobius algebra.

This singularity is however not the only one. A full geometrical description is somewhat involved and we simply content ourselves with observing the dimension, as summarised in Table~\ref{Singe}.
\begin{table}[h]
\[
\begin{array}{|c|c||c|c|}\hline
\multicolumn{2}{|c||}{\mbox{$\cV_\varphi$ with $n=2$}}&\multicolumn{2}{c|}{\mbox{$\cV_\varphi^*$ with $n=3$}}\\  \hline
 N & \mbox{singular locus dimension} & N & \mbox{singular locus dimension} \\ \hline \hline
 4&1&6&0\quad\mbox{(point at the origin)}\\ \hline
 5&2&7&1 \\ \hline
 6&2&8&1 \\ \hline
\end{array}
\]
{\caption{\label{Singe} {\sf Affine dimension of the singular locus of the corresponding $3$-dimensional varieties}.}}
\end{table}

We see that the constraints~\eqref{HWfix} on the tensor coefficients $\lambda$ for $\cV^*$, reminiscent from the highest weight conditions, also implies the existence of singularities. However, the nature of such singularities is different from the corresponding $3$-dimensional varieties $\cV_\varphi$ as they differ in their dimension and produce different topological numbers, such as the Euler characteristic $\chi$ and the Hodge diamond.

%%%%%%%%%%%%%%%%%%%%%%%%%%%%%%%%%%%%%%
\section{Discussion and Outlook}

In this work, we presented geometrical properties of algebraic varieties built up from irreducible representations of the \Mobius algebra in the hope that it will help to shed light on their physical meaning and, as a consequence, on the understanding of the scattering equation.

We found that these varieties are all {\it affine Calabi-Yau} manifolds with Hilbert series following very regular pattern, such as the Mahonian triangles or the cyclotomic polynomials. For the physical \Mobius invariant three-dimensional varieties, we furthermore found that they consist of familiar geometries such as the {\it conifold} for the scattering of four particles, a cone over a {\it K3} surface for five particles and a cone over {\it Fano} surfaces (i.e., del Pezzo surfaces) for six and seven particles. 

In addition, computation of the Euler number and Hodge diamonds showed that not only are the affine Calabi-Yau spaces singular, but so too are their projectivizations to compact surfaces. A short computation unveiled that the singular points are related to the conditions for the polynomial system to be an irreducible representation of the \Mobius algebra. The singularity contains at least one point for the projective varieties and it would be expedient to understand its physical meaning.
Indeed, the physical constraints on the momenta and hence coefficients in the scattering variety show that we are in very special singular points in the complex structure moduli space.

It would be interesting to develop further the geometrical properties, such as Hodge numbers for more varieties, and a complete description of the singularity geometry. Furthermore, our study focused on generic parameters and very special momenta configurations might lead to extra geometrical structures. It would be interesting to investigate the way the varieties degenerate when one (or more) of the polynomial coefficient $\lambda_S$ vanishes and understand the dependence of the geometrical properties of the varieties on the structure of these coefficients.
Understanding the physical meaning of the above varieties is of crucial importance and we hope that our analyses for the geometrical structure of the scattering variety offer a good starting point to unveil deeper connections.

%%%%%%%%%%%%%%%%
\section*{Acknowledgements}
YHH would like to thank the Science and Technology Facilities Council, UK, for an Advanced Fellowship and for STFC grant ST/J00037X/1, the Chinese Ministry of Education, for a Chang-Jiang Chair Professorship at NanKai University, the city of Tian-Jin for a Qian-Ren Scholarship, the US NSF for grant CCF-1048082, as well as City University, London, the Department of Theoretical Physics and Merton College, Oxford, for their enduring support. CM is grateful to Hwasung Lee for helpful discussions as well as City University, London and Helios Technology for giving the possibility to pursue this work.

\newcommand\href[1]{\url{#1}}
%%%%%%%%%%%%%%%%%%%======================================================
%%%%%%%%%%%%%%%%%%%======================================================

\end{document}